 \documentclass[12pt,preprint]{aastex}

 \input{psfig}

\def\gax    {${_>\atop^{\sim}}$}
\def\aox    {$\alpha_{ox}$}
\def\etal   {{\it et al.}~}
\def\lya    {Ly$\alpha$~}

\def\nh     {N$_{\rm H}$~}

\def\chandra {{\it Chandra}~}
\def\sdssone   {SDSSP J083643+005453~}
\def\sdsstwo   {SDSSP J103027+052455~}
\def\sdssfour   {SDSSP J130608+035626~}
\def\sdssthree   {SDSSP J104433-012502~}

\shorttitle{Chandra detections of z$\sim$ quasars}
\shortauthors{Mathur \etal}

\begin{document}


\title{\chandra Detection of Highest Redshift (z$\sim$6) Quasars in X-rays}


\author{Smita Mathur\altaffilmark{1,2}, 
Belinda J. Wilkes\altaffilmark{2} \&  
Himel Ghosh\altaffilmark{2} 
}





\altaffiltext{1}{The Ohio State University; smita@astronomy.ohio-state.edu}
\altaffiltext{2}{Harvard Smithsonian Center for Astrophysics; belinda, 
hghosh@head-cfa.harvard.edu}


\begin{abstract}
We report on \chandra observations of three quasars \sdssone, \sdsstwo
and \sdssfour at redshifts 5.82, 6.28 and 5.99 respectively. All the
three sources are clearly detected in the X-ray band, up to rest frame
energies of $\sim 55$ keV. These observations demonstrate the
unprecedented sensitivity of \chandra to detect faint sources in
relatively short exposure times (5.7--8.2 ksec). The broad band X-ray
properties of these highest redshift quasars do not appear to be any
different from their lower redshift cousins. Spectra of the sources
could not be determined with only few counts detected. Observations
with XMM-Newton will be able to constrain the spectral shapes, if they
are simple. Determination of complex spectra in a reasonable amount of
time, however, will have to await next generation of X-ray missions.
\end{abstract}




\section{Introduction}

The highest redshift (z) quasars are of interest not only for their
`record setting' quality, but also because they can tell us about the
formation of quasars and about conditions in the first few percent of
the age of the universe. The discovery of z$\sim 6$ quasars (Fan \etal
2001; hereafter Paper I) was remarkable as it showed that luminous
quasars with massive black-holes had formed when the Universe was less
than a Gyr old. Quasars \sdssone, \sdsstwo, \sdssthree and \sdssfour
at z=5.82, 6.28, 5.80 and 5.99 respectively were discovered as a part
of the Sloan digital sky survey (SDSS, York \etal 2000) imaging
multicolor observations. Follow-up spectroscopy revealed rich quasar
spectra with strong \lya forest. The importance of these observations
was amplified by the realization that the neutral hydrogen fraction of
the intergalactic medium increases drastically around z$\sim 6$,
marking it to be the epoch of end of reionization (Becker \etal 2001,
Fan \etal 2002, Pentericci \etal 2002).

Observations of high redshift quasars are also important to understand
quasar evolution and associated, possible, growth of massive black
holes. While the evidence for the evolution of the quasar luminosity
function is clear (Boyle \etal 2000), we still do not know {\it how}
the quasars evolve, i.e. which of observed properties of quasars
exhibit change with time. Mathur (2000) suggested that accretion rate
relative Eddington is likely to be a function of time, in which case
properties related to accretion, such as those contributing to
``eigenvector 1'' (Boroson 2001) would be expected to change.  X-ray
studies of high redshift quasars are important as the X-ray slope
changes with accretion rate and so may give us information about the
evolution. The overall spectral energy distribution of quasars also
changes with accretion rate. While small changes are difficult to see,
the extreme cases, e.g. advection driven accretion flows (ADAF) show
clearly different spectral energy distributions, including that in the
X-ray band (Narayan \etal 1998 and references there in). So the hope
of X-ray studies of high redshift quasars is that they may give us
some clue about early accretion process, and so, possibly, about their
evolution.

This goal has not been achieved yet, because X-ray detections of high
redshift quasars have been rare. Prior to \chandra and XMM-Newton era,
only a handful of z$>4$ quasars were detected in X-rays (Kaspi \etal
2000 and references there in). Thanks to the sensitivity of the
\chandra observatory, the total number of X-ray detected z$>4$ quasars
is now over two dozen (Silverman \etal 2002, Vignali \etal 2001,
hereafter Paper II, and references there in). At redshift $>5$,
however, the numbers are very sparse. The z=5.8 quasar \sdssthree was
detected with XMM-Newton (Brandt \etal 2001, Mathur 2001) and a z$\sim
5.2$ quasar is reported to be detected with \chandra (Brandt \etal
2002) (A z=5.27 quasar SDSSp J120823.82+001027.7 is also reported to
be detected by Vignali \etal 2001, but with only two counts, this is
questionable).  In spite of these small numbers, there have been
tantalizing suggestions of changes of X-ray properties of quasars with
redshift (e.g. Paper II, and references there in). Vignali \etal
(2001) found suggestive evidence that \chandra detected high redshift
quasars are fainter in X-rays compared to those detected earlier with
ROSAT. Mathur (2001) found that the z=5.8 quasar \sdssthree is likely
to have a steep X-ray slope. While these suggestions need to be
confirmed with better observations, trends are encouraging.  In this
paper we report the \chandra\ detections of three z$\sim 6$ SDSS
quasars.

\section{Observations and Data Analysis}

The sub-arcsecond angular resolution and very low background makes
\chandra ideal for detecting faint sources like high redshift quasars.
The z$\sim 6$ quasars \sdssone, \sdsstwo, and \sdssfour were observed
with \chandra X-ray observatory on 29 January 2002 with director's
discretionary time. The data were made available to the astronomical
community immediately. We downloaded the data from the \chandra X-ray
Center (CXC) archive. Observation details are given in Table 1.

All the three observations were performed using the detector Advanced
CCD Imaging Spectrometer for spectroscopy (ACIS-S). The sources were
observed at the nominal aim-point on the back illuminated CCD
S3. Analysis was performed in a standard manner using the software
\chandra\ interactive analysis of observations (CIAO version 2.2.1;
Elvis, in preparation) \\ (see also \chandra data processing threads at
http://cxc.harvard.edu/ciao/documents\_threads.html).

Virtually all source counts are contained in the 2.9\arcsec\ (5.75
pixel) radius extraction aperture that was centered upon the X-ray
source centroid.  Background counts were estimated from an annulus
with inner and outer radii of 10 and 30 pixels, respectively. The
background counts expected in the source extraction region are between
0.2 and 0.3 for the three sources, so we ignore the background in rest
of the analysis. All the three sources are clearly detected in the
broad band (0.5--8.0 keV). To determine the centroid position of the
sources we used CIAO tool {\bf wavdetect} (Freeman \etal 2002). The
X-ray positions of all the three sources are within $1^{\prime\prime}$
of the optical positions (Table 2).  We also extracted source counts
in the soft (0.5--2.5 keV) and hard (5--8 keV) bands separately.
Figure 1 shows the broad band ACIS-S images.

We calculated the probability of serendipitous source detection using
the observed source flux distribution of Hasinger \etal
(1998). Assuming the 0.5--2.5 keV flux limit of $1\times 10^{-15}$
ergs cm$^{-2}$ s$^{-1}$ for the faintest of the three sources, and a
detection cell of $2.9^{\prime\prime}$ radius, the expected number of
sources is $2.4\times 10^{-3}$ for each of the observation. So it is
highly unlikely that the detected sources are unrelated to the
quasars. We therefore conclude that SDSS sources observed by \chandra
are detected.

Using the observed count rate, the broad band flux was determined
assuming a power law slope of $\alpha=1.0$ and the Galactic \nh\ values
for each source (Table 2).  The rest frame luminosity was calculated for
two different cosmologies: H$_0=50$ km s$^{-1}$ Mpc$^{-1}$ and
q$_0=0.5$ for consistency with earlier work and also with H$_0=75$ km
s$^{-1}$ Mpc$^{-1}$ and q$_0=0.5$. The observed properties of all the
three sources are presented in Table 2.

\section{Discussion}

The detections of z$\sim 6$ quasars in X-rays demonstrates the power
of \chandra to detect such faint sources. The sources are detected up
to rest frame energies of $\sim 55$ keV. With only a few counts
detected, however, detailed source information is difficult to
obtain. Nevertheless, we can learn about the overall broad band
spectral energy distributions by determining the optical to X-ray
ratio \aox\ (defined as $-log(f_{X}/f_{opt})/log(\nu_{X}/\nu_{opt})$
where $f_{X}$ and $f_{opt}$ are monochromatic flux densities
 at 2 keV and 2500 \AA\ respectively in the rest frame of a
quasar; Tananbaum \etal 1979). We calculated the optical monochromatic
luminosity at 2500 \AA\ using observed optical flux (Paper I) and
assuming a power-law optical continuum slope with $\alpha=0.5$. The
use of $\alpha=0.5$ is appropriate for the observed continuum in Paper
I. The monochromatic luminosity at 2 keV was calculated assuming
power-law slope $\alpha=1$. The resulting \aox\ values are 1.49, 1.61,
and 1.48 for \sdssone, \sdsstwo and \sdssfour respectively (Table
2). These are within the range observed for normal lower redshift
quasars (e.g. Brandt \etal 2001). Thus the broad band X-ray properties
of these highest redshift quasars do not appear to be any different
from lower redshift objects.

For comparison with earlier work (Silverman \etal 2002, Paper II), in
Figure 2 we have plotted the observed X-ray fluxes of z$>4$ quasars
vs. their AB$_{1450(1+z)}$ magnitudes along with those reported here.
The z$\sim 6$ quasars do not occupy any conspicuously different region
on this plot, nor do the X-ray selected (ROSAT) sources look
particularly bright (\S 1).  In Figure 3, we have plotted the X-ray to
optical flux ratio as a function of redshift. No obvious trend with
redshift is apparent.

X-ray astronomers have traditionally used hardness ratios to extract
spectral information when observed counts are too few to perform
spectral analysis (Mathur 2001 and references there in). The hardness
ratio is defined as HR=(soft $-$ hard)$/$(soft $+$ hard) where
``soft'' and ``hard'' are the counts in energy bands 0.5$-$2.5 keV and
2.5$-$8.0 keV respectively (we prefer HR over the band ratio
BR=soft$/$hard sometimes used in the literature. HR is bounded well
between $-1$ and 1 for detected sources, compared to the 0 to $\infty$
range of BR). In these high-redshift sources we are observing rest
frame energies above $\sim 3.5$ keV so that only highly absorbed
sources would show evidence for absorption and we expect to be
observing the pure continuum emission.  The HR ratio of \sdssone is
consistent with that expected with a power-law slope $\alpha=1$. This
object, however, is radio-loud (Paper I) and so its X-ray slope is
expected to be flatter around a mean $\alpha=0.5$ (Wilkes \& Elvis
1987). For \sdsstwo, no counts are detected in the hard band,
resulting in HR=1. A power-law with slope $\alpha$\gax 2 would then be
required. For \sdssfour, on the other hand, a flatter power-law with
$\alpha \sim 0.3$ would be required to reproduce the observed HR of
0.56. We caution, however, against attaching too much significance to
the above results, given the errors associated with observed counts in
each band. Within the errors, the sources are consistent with the
normal power-law slope expected for quasars, except possibly for
\sdsstwo.

Possibly, the most remarkable thing about these highest redshift
quasars is that they are so absolutely unremarkable in both optical
and X-ray bands implying no strong evolution out to redshifts $\sim
6$.  The only possible difference could be in the X-ray spectral slope
which could not be determined with the \chandra data.  A lot more
counts are needed to determine the exact spectral shape of these
objects. The expected count rates with XMM-Newton are 11, 3 and 6 per
ksec for the three objects respectively, for combined EPIC-pn and
EPIC-MOS detectors with thin filters. For the brightest two objects a
crude spectrum with about 300 counts can be obtained in a reasonable
observing time of 25--50 ksec. We have an on-going program with
XMM-Newton to do precision spectroscopy of some z$> 4$
quasars. Extending this program to highest redshifts would be
particularly interesting. Clearly, we have to work harder and look
deeper to understand quasar evolution.

\acknowledgments We thank the entire \chandra team for a very
successful mission. We gratefully acknowledge support from grant
GO-2118X from Smithsonian Astrophysical Observatory to SM and NASA
contract NAS8-39073 (CXC) to BJW and HG.

\clearpage
\thispagestyle{empty}


\newpage
\begin{table}[h]
\caption{\chandra observations of z$\sim 6$ quasars}
\begin{tabular}{|lcccccc|}
\tableline
Source & RA   & Dec   & z  & Obs. date & Sequence \# & Exposure 
 \\
&&&&&& (s)\\
\tableline
SDSSP J0836+0054 & 08 36 43.90 & +00 54 53.30 & 5.82 & 2002/1/29 & 700605 & 5687.06
\\
&&&&&& \\
SDSSP J1030+0524 & 10 30 27.10  & +05 24 55.00 & 6.28 & 2002/1/29 & 700606 & 7955.57
\\
&&&&&& \\
SDSSP J1306+0356 & 13 06 08.30 & +03 56 26.30 & 5.99 & 2002/1/29 & 700607 & 8157.78 \\
\tableline
\end{tabular}
\smallskip
\end{table}

\newpage
\begin{table}[h]
\caption{Properties of z$\sim 6$ quasars}
\begin{tabular}{|lccccccccc|}
\tableline
Source & \multicolumn{2}{c}{X-ray Position} & N$_H$$^a$ & 
\multicolumn{3}{c}{Detected Counts} & f$_X$$^b$ & L$_X$$^c$ & 
$\alpha_{OX}$ \\
       &   RA & DEC & & Broad & Soft  & Hard & & &\\  
\tableline
SDSSP J0836+0054 & 08 36 43.9 & +00 54 52.9 & 4.12 & 21$\pm 4.6$ & 19$\pm 4.4$ & 2$\pm 
1.4$ & 2 & 6.1 & 1.49\\
&&&&&&&& 2.7 & \\
&&&&&&&&& \\
SDSSP J1030+0524 & 10:30:27.1 & +05 24 54.7 & 3.12 & 6$\pm 2.4$ & 6$\pm 2.4$ & 0 & 0.4 
& 1.4 & 1.61\\
&&&&&&&& 0.64 & \\
&&&&&&&&& \\
SDSSP J1306+0356 & 13:06:08.3 & +03 56 26.9 & 2.07 & 18$\pm 4.2$ & 14$\pm 3.7$ & 
4$\pm 2$ & 1 & 3.5 & 1.48\\
&&&&&&&& 1.5 & \\
&&&&&&&&& \\
\tableline
\end{tabular}
\smallskip
\smallskip
a:  Galactic column density calculated using HEASARC tool {\bf nh}, in the 
units of 10$^{20}$ atoms cm$^{-2}$ \\
b: Broad band flux corrected for Galactic absorption, in the units of
10$^{-14}$ erg s$^{-1}$ cm$^{-2}$  \\
c: Luminosity in the rest frame 0.5--8 keV band, in the units of 10$^{45}$ erg 
s$^{-1}$. For each source, the
first line corresponds to H$_0=50$ km s$^{-1}$ Mpc$^{-1}$ and
q$_0=0.5$, while the second line assumes H$_0=75$ km s$^{-1}$
Mpc$^{-1}$ and q$_0=0.5$. \\
\end{table}



\clearpage

\noindent
\figcaption{ACIS-S images of the SDSS sources, smoothed by two
pixels. Each image is 20$^{\prime\prime}$ $\times$ 20$^{\prime\prime}$
and the circle is centered on the X-ray position with radius
2.9$^{\prime\prime}$. Top: \sdssone, middle: \sdsstwo,
bottom: \sdssfour}

\noindent
\figcaption{For comparison with earlier work, observed, Galactic
absorption corrected X-ray flux is plotted against
AB$_{1450\times(1+z)}$ for z$>4$ quasars. Only detected sources are
plotted. Data are from Paper II, Silverman \etal 2002, and this
paper. Optical magnitudes of the SDSS quasars are from Paper I. The
z$\sim 6$ quasars do not occupy any conspicuously different region on
this plot}

\noindent
\figcaption{X-ray to optical flux ratio plotted as a function of
redshift. Symbals are as in Figure 2. Optical f$_{\nu}$ is calculated from
published AB$_{1450\times(1+z)}$ magitudes following Fukugita \etal
(1996). The dashed line corresponds to \aox=1.6 at z=0. It is clear that
the redshift 6 quasars are not very different from their lower
redshift cousins.}

\newpage
\clearpage
\begin{figure}[h]
\psfig{file=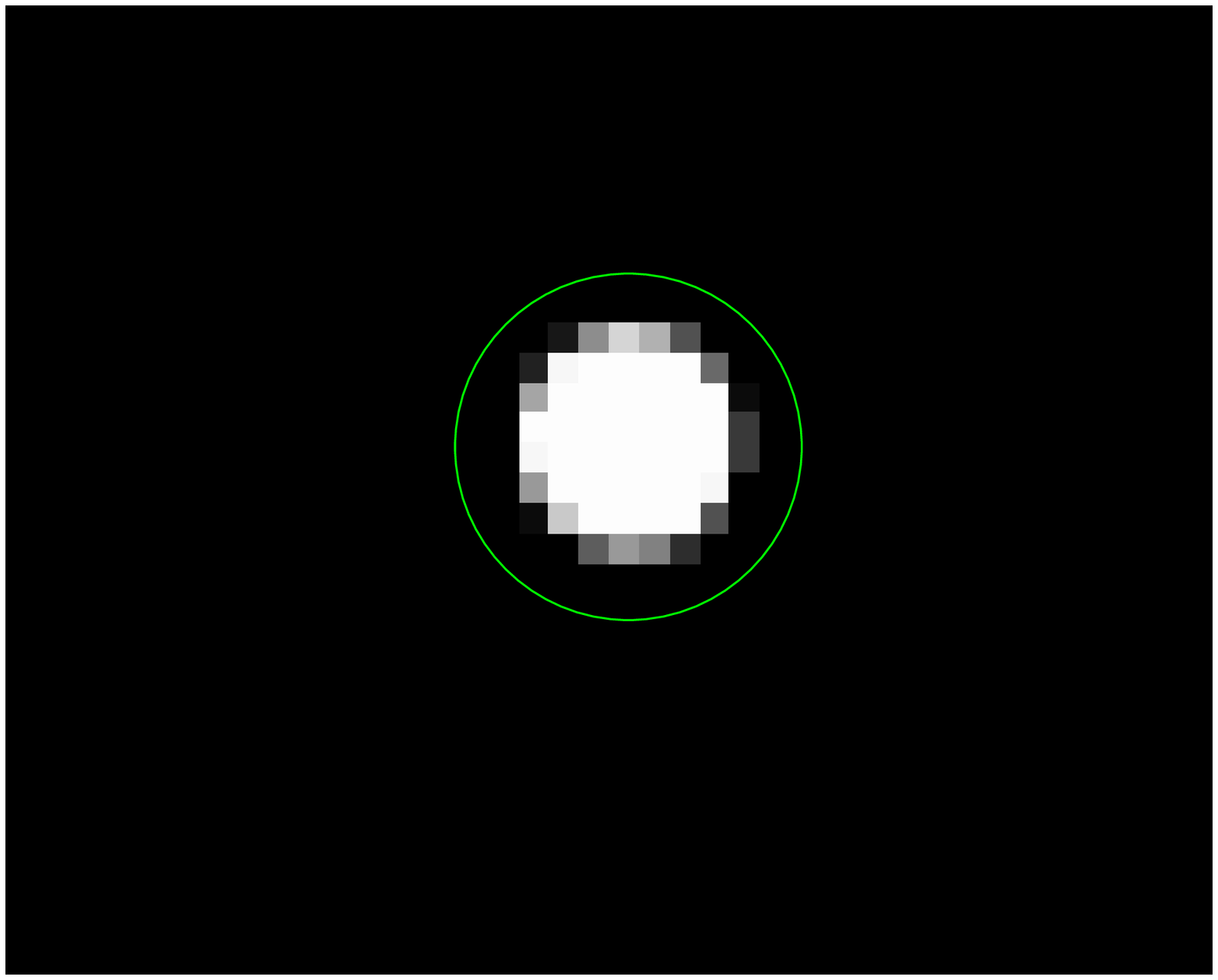,height=3.in}
\psfig{file=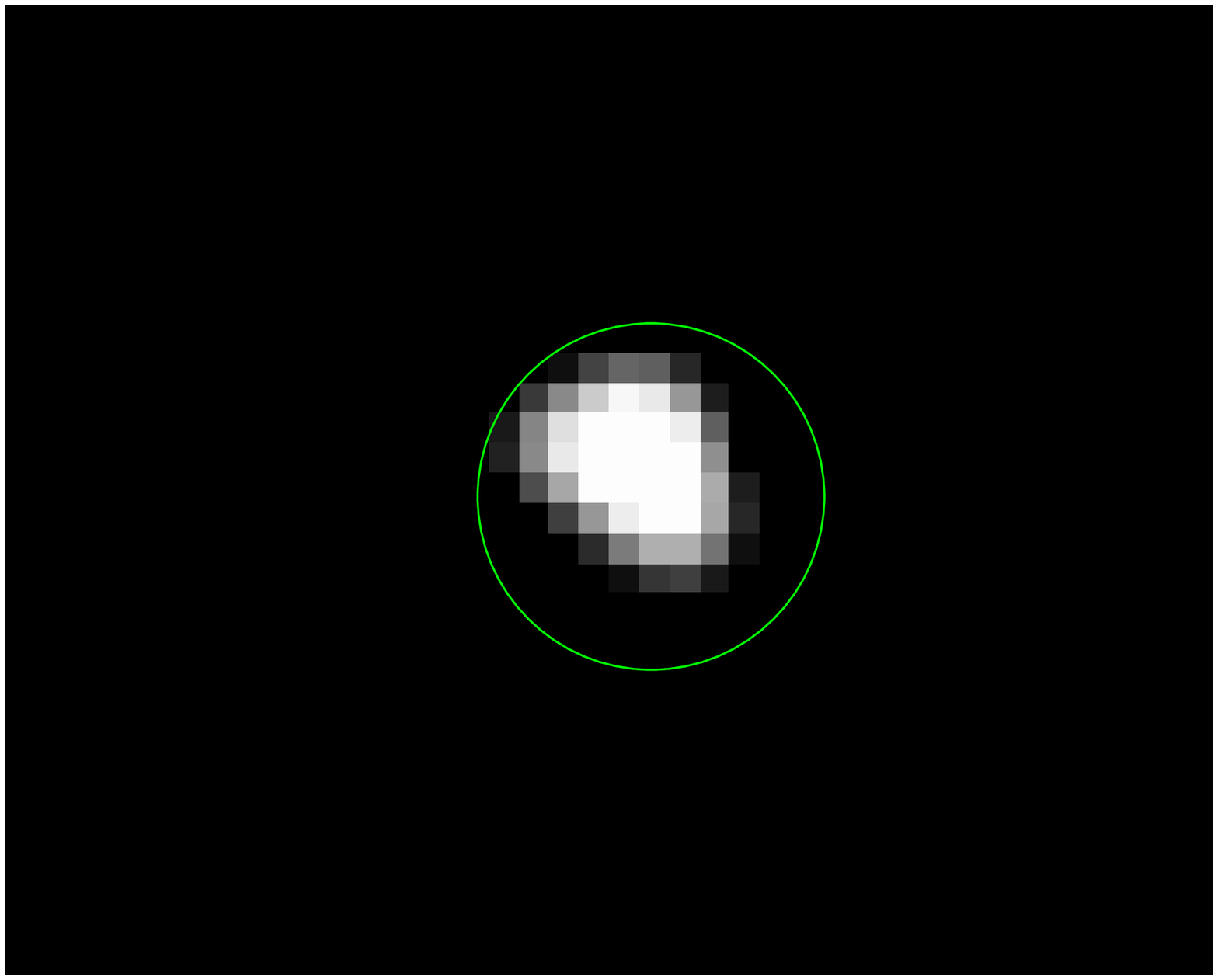,height=3.in}
\psfig{file=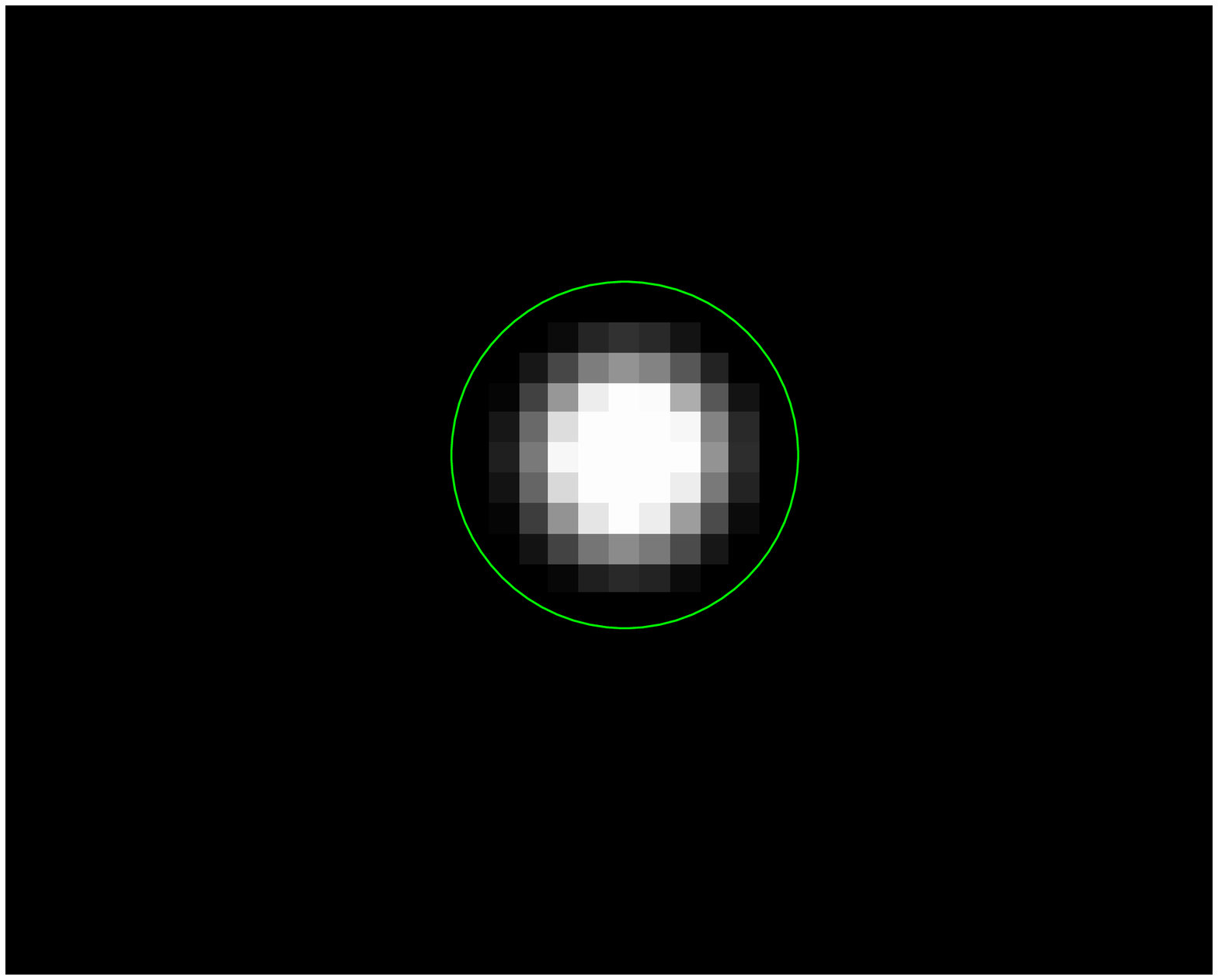,height=3.in}
\end{figure}
\clearpage

\newpage
\clearpage
\begin{figure} [h]
\psfig{file=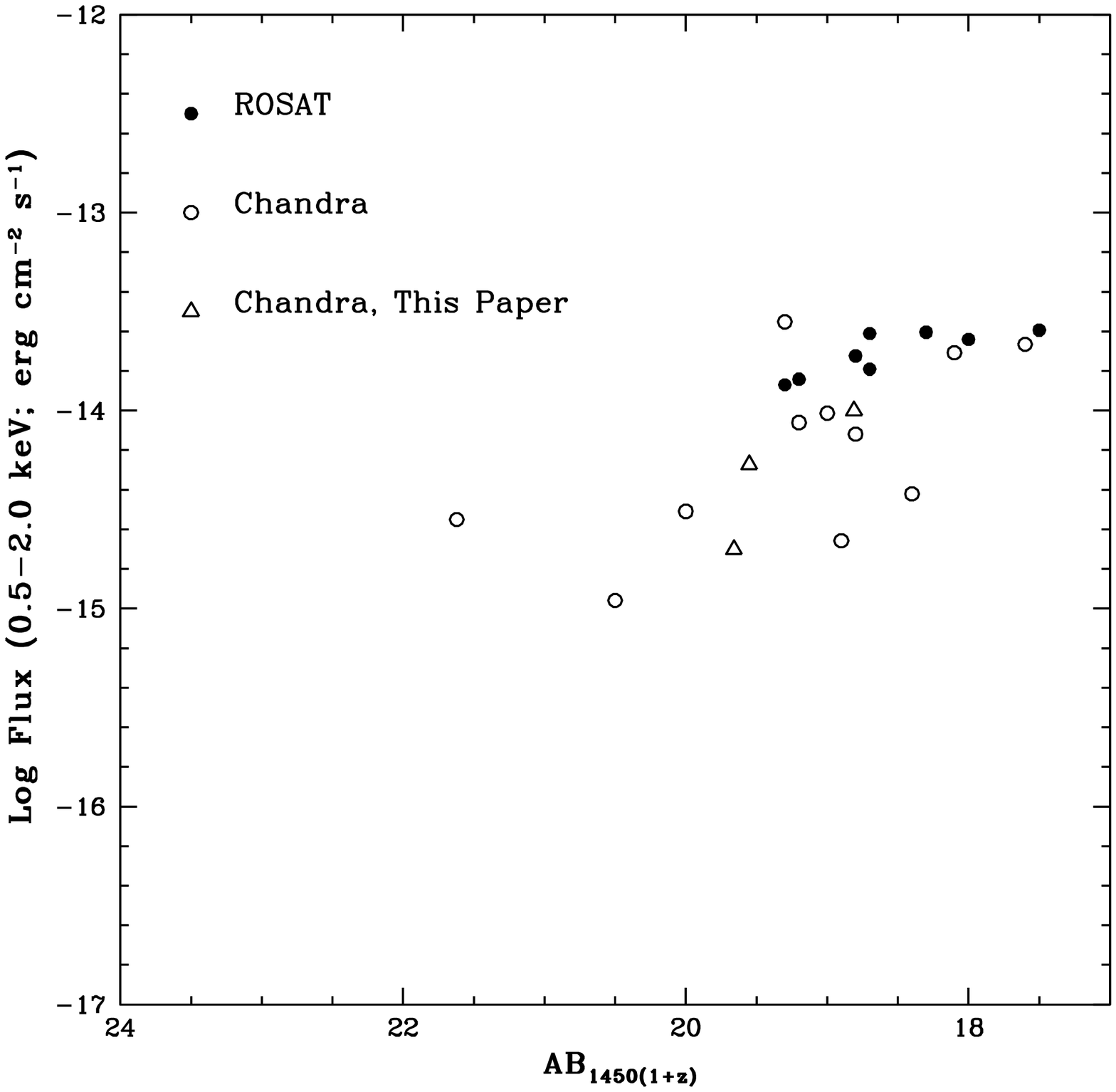,height=6.in,width=6.0in}
\end{figure}
\clearpage

\newpage
\clearpage
\begin{figure} [h]
\psfig{file=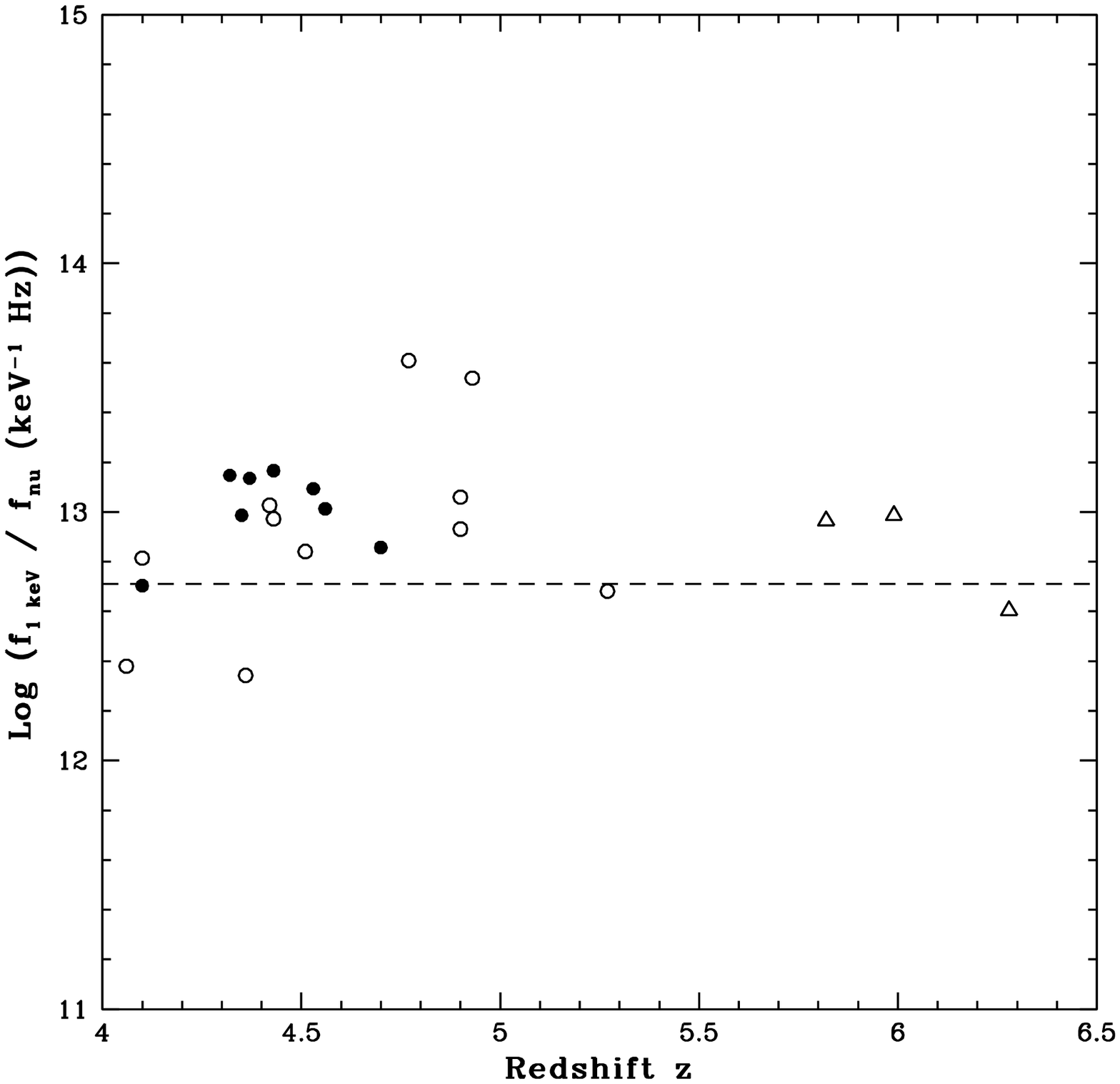,height=6.in,width=6.0in}
\end{figure}
\clearpage


\end{document}